\documentclass[aps,prc,twocolumn,showpacs,10pt]{revtex4-1}

\usepackage[utf8]{inputenc}
\usepackage{amsfonts}
\usepackage{amsmath}
\usepackage{amssymb}
\usepackage{epsfig}
\usepackage{bm}

\begin{document}

%\preprint{}

\title{Direct and sequential  four-body recombination rates at  low temperatures}



%\thanks{}

\author{E. Garrido}
\email{e.garrido@csic.es}
\affiliation{Instituto de Estructura de la Materia, IEM-CSIC, Serrano 123, E-28006 Madrid, Spain}
\author{A.S. Jensen}
\affiliation{Department of Physics and Astronomy, Aarhus University, DK-8000 Aarhus C, Denmark}


\date{\today}

\begin{abstract}
  We investigate four-body nuclear reactions in stellar
  environments contributing to creation of light nuclei, exemplified
  by $^9$Be and $^{12}$C. The originally assumed process is radiative
  capture, where nuclear clusters combine into the excited final
  nucleus and photon emission populates the stable nuclear ground
  states.  Instead, we consider nuclear four-body recombination
  reactions where a spectator nuclear particle replaces the photon.
  We first develop the elaborate formalism for both, direct and
  sequential capture processes, where the decaying three-body
  resonance is formed without and with population of an intermediate
  two-body resonance, respectively.  To facilitate both calculations
  and practical applications we parameterize the involved cross sections 
   as done successfully in previous computations
  of reaction rates.  We consider the lowest-lying nuclear states with
  their dominant contributions at low stellar temperatures.  We
  calculate and compare reaction and production rates for different
  processes.  The direct reaction mechanism dominates by many orders
  of magnitude at low temperature, where the sequential stepping
  stones are energetically too expensive to use. At somewhat higher
  temperatures these two different nuclear four-body mechanisms become
  comparable.  Comparison to radiative three-body capture reveals
  already formally, but also numerically, that four-body nuclear
  recombination must dominate for sufficiently high nuclear densities.
  Numerical values are given for all these rates as function of
  temperature and density.  The relative importance are exhibited.
\end{abstract}

%\pacs{ }%21.45.-v, 26.20.-f, 26.30.-k,27.20.+n}
%\keywords{}

\maketitle

\section{Introduction}

The spontaneous radiative capture of three alpha particles to produce $^{12}$C is known to play a crucial role in order to
bridge the $A=5$ and $A=8$ gaps in the nuclear chart \cite{apr05}. This is a three-body reaction that becomes relevant
in high density scenarios due to the absence of alternative simpler routes to the $^{12}$C production.
This process is usually treated assuming a sequential mechanism \cite{ang99}, i.e., as two consecutive two-body processes, where in a first step 
the $0^+$ particle-unstable state in $^8$Be is populated. The small width of this resonance permits, in the second step, to capture a third
alpha particle and form $^{12}$C before decaying of the $^8$Be resonance. 
The width of the two-body resonance, even if very small, is still playing an important role, as argued in \cite{nom85,lan86}, where it  is shown
that  for sufficiently low temperatures ($T \lesssim 3\times 10^7 \mbox{ K}$) the capture proceeds through the low energy wing
of the alpha-alpha resonance, which gives rise to a significant increase of the reaction rate.
Furthermore, in \cite{gar11} it has been shown that a direct
capture mechanism, without population of intermediate two-body states, could also be crucial at very low temperatures,
where the particle energy is still too small to populate the intermediate state. The direct mechanism could actually be responsible for
an increase of several orders of magnitude of the reaction rate in the low-temperature region. 

Together with radiative capture processes, it has been suggested in \cite{tru69} that pure nuclear interactions could also play an important role in the de-excitation 
of $^{12}$C excited states. Inelastic collisions with surrounding neutrons, protons, alpha particles, and perhaps other ions,
were estimated to substantially increase the rate of formation of $^{12}$C for neutron densities above $10^7$ g/cm$^3$
and temperatures greater than a few GK. Preliminary studies of the reaction rate enhancement produced by inelastic
alpha and proton scattering were reported in \cite{mor70,dav71}.

Recent calculations  have corroborated these conclusions. In \cite{die10,die14} the possibility of four-body recombination, or
particle induced de-excitation (both denominations will be indistinctly used in this work), was pointed out as an alternative mechanism to the spontaneous three-body
radiative decay. For densities higher than $10^6$ g/cm$^3$ the four-body recombination 
dominates, especially for temperatures below 1 GK.

This sort of particle-induced reaction rate enhancement has recently been revisited \cite{bea17}. The theoretical 
background behind this work is as in \cite{tru69,mor70,dav71}, but considering also the possibility of the neutron
as the particle that boosts the decay. The main difference is that in \cite{bea17} a more sophisticated procedure in 
order to obtain the crucial alpha, proton, and neutron inelastic cross sections on $^{12}$C has been employed. Again, a relevant
enhancement of the reaction rate was found for sufficiently large temperatures and densities.

In these works \cite{tru69,mor70,dav71}, and also in the recent paper \cite{bea17}, the reaction rate was obtained by 
describing the process as a simple two-body process, where a previously populated three-body resonance, due
to the presence in the environment of neutrons, protons, or alpha particles, decays into a three-body bound state, such that
the excess of energy is transformed into kinetic energy of the fourth particle. Thanks to the detailed balance principle, 
this process can be easily related to the inverse reaction, namely the inelastic excitation of the bound three-body
state by the fourth particle, either a neutron, or a proton, or an alpha particle. The available experimental
information or the reliability of the theoretical calculation of these inelastic cross sections becomes then
crucial in order to obtain also reliable estimates of the reaction rates.

This strong simplification (to a two-body process) of what really is a four-body process amounts to ignore the fact that the decaying three-body
resonance actually has a finite width, and therefore a finite lifetime, treating  in practice the resonance as a bound state with 
an energy equal to the resonance energy. However, it is known that for radiative decay the finite width of the resonance 
determines to a very large extent the value of the reaction rate, especially a low temperatures \cite{ang99}. Furthermore, 
not only the width, but also the reaction mechanism, through which the resonance is populated, plays an important role. In 
\cite{gar11} it is shown how, for temperatures about 0.01 GK, a direct population of the resonance could lead to radiative reaction rates, 
for formation of $^{12}$C or $^9$Be, of up to seven orders of magnitude bigger than by the sequential mechanism, where the 
intermediate low-lying $0^+$ resonance in $^8$Be is populated before capturing the third particle.

These two elements, the finite resonance width, and the reaction mechanism, should also play an important role
in the case of four-body recombination reactions, and they could therefore also be relevant when comparing to 
the radiative process and the circumstances under which the particle-induced reaction rate is dominating.
The main goal of this work is therefore to investigate the effect of these two ingredients, resonance width and
reaction mechanism, in the four-body recombination reaction rates, and check how these rates compete
with the ones corresponding to a spontaneous radiative process. The cases of formation of $^9$Be and
$^{12}$C induced by the presence of protons, neutrons, or alpha particles will be considered.

This work follows a scheme similar to the one in Ref.~\cite{gar11}, where the radiative capture process leading
to $^9$Be and $^{12}$C was investigated. We first give the general expressions for the four-body
recombination reaction rate. The analytic expressions corresponding to a direct and a sequential
capture mechanism are derived in Sections III and IV, respectively. For each of these two cases,
the expressions for the limit of zero-width for the resonances are also derived. 
In order to facilitate the reading of the paper and avoid unnecessary rereading of Sections III and IV,
we  summarize in Section V the key equations relevant for the subsequent calculations
assuming either direct or sequential capture mechanism. Section V is closed with some remarks concerning the production rates and
the comparison with the rate corresponding to the radiative capture process. The numerical results are presented in details in 
Section VI for the important cases of $^9$Be and $^{12}$C. We close the paper with the summary and conclusions
presented in Section VII.

\section{The four-body nuclear recombination rate}

The goal in this work  is to investigate the four-body recombination rate for a given reaction 
$a+b+c+d \rightarrow A+d$, where $A$ is a bound state formed by particles $a$, $b$, and
$c$. In general, particles $a$, $b$, and $c$ will be assumed to populate a three-body 
resonance, which will be  in fact the one decaying into the bound state $A$. Particle $d$ is simply  
acting as a mere spectator, and its only role is to maintain energy and momentum in the 
resonance decay process. Reactions where particle $d$ is exchanged with one of the ones 
forming the three-body resonance will not be considered here.

Let us denote the initial four-body energy as $E$. Therefore, by energy conservation we must
have that $E=T'_{Ad}+B=T'_{Ad}-|B|$, where $B$ is the binding energy of the three-body system $A$, 
and $T'_{Ad}$ is the final relative kinetic energy between $A$ and $d$. In Ref.~\cite{die14} it is shown
that the general expression for the four-body reaction rate, averaged over the four-body
Maxwellian energy distribution, takes the following form as a function of the temperature $T$:
\begin{eqnarray}
\lefteqn{\hspace*{-1cm}
\langle R_{abcd}(E) \rangle= \frac{4 (2\pi)^{\frac{5}{2}} \hbar^6 \mu_{A,d}}{(\mu_{a,b} \mu_{ab,c} \mu_{abc,d})^{3/2}}
\frac{g_A}{g_a g_b g_c} } \nonumber \\ && 
\frac{1}{(K T)^{\frac{9}{2}}}
\int_0^{\infty} T'_{Ad} \sigma_{Ad}(T'_{Ad}) e^{-\frac{E}{K T}} dE \; ,
\label{4bdrr}
\end{eqnarray}
where $K$ is the Boltzmann constant,  $\sigma_{Ad}$ is the cross section for the breakup reaction 
$A+d \rightarrow a+b+c+d$, and $\mu_{i,j}$ represents the reduced mass
of particle $j$ and particle (or group of particles) $i$.

The cross section $\sigma_{Ad}$ is related to the capture
cross section $\sigma_{abcd}$ of the inverse process by means of the principle of
detailed balance.  Both cross sections contain the same matrix element, $\langle \Psi_i |W | \Psi_f\rangle$,
where $\Psi_i$ and $\Psi_f$ are the initial and final four-body wave functions, and
$W$ represents the interaction responsible for the reaction. Therefore, calculation of the
reaction rate in Eq.(\ref{4bdrr}) requires in principle knowledge of the two four-body wave functions, 
one of them representing the four particles in the continuum, and the other one describing a bound
three-body state plus the fourth particle in the continuum.

Therefore, calculation of Eq.(\ref{4bdrr}) is in general a rather complicated task, which,
unless some simplifications, like in Ref.~\cite{die14}, are made, unavoidably requires calculation of
four-body wave functions. Furthermore,  even if some simplifications are introduced, the
competition between the direct and sequential mechanism is hidden by the procedure itself. It would be rather difficult to extract from the structure of the four-body states which of the two
mechanisms has actually been predominantly used.  

For this reason we consider it useful to perform the 
same kind of analysis for the four-body recombination rate as made in Ref.~\cite{gar11} for the three-body 
radiative capture process. The idea is to compare  the direct and sequential capture mechanisms 
in the four-body recombination reaction rates using realistic phenomenological parametrizations 
of the cross-sections entering in the reactions. In this way it will be possible to study in a clear way the
effects arising from the different positions and widths of the  two- and three-body resonances   
involved, as well as the effects produced by these two capture mechanisms.

As we will see, the analytical expressions describing the process under investigation will contain
two different types of cross sections. The first type corresponds to capture processes leading to 
a two-body or a three-body resonance, which will be described as simple Breit-Wigner shaped 
cross sections, depending on the energy and width of the populated resonance.
The second type is the cross section $\sigma_{Ad}$, present also in Eq.~(\ref{4bdrr}),
but now understood as an inelastic two-body process where the system $A$ is just excited
after the collision with particle $d$.
Following \cite{ang99,fow67}, when $A$ and $d$ are both charged, 
the inelastic cross section takes the general form:
\begin{equation}
\sigma_{Ad}(T'_{Ad})=\frac{S(T'_{Ad})}{T'_{Ad}} \exp \left(-2\pi \eta \right),
\label{sad}
\end{equation}
with $\eta=Z_A Z_d e^2/(\hbar v)$, where $e$ is the electron charge, $Z_A$ and $Z_d$ are the charges of the two
particles,  and $v$ their relative velocity. If either $A$ or $d$ is not charged we take 
instead \cite{fow67}:
\begin{equation}
\sigma_{Ad}(T'_{Ad})=\frac{S(T'_{Ad})}{\sqrt{T'_{Ad}}}.
\label{sad2}
\end{equation}

The $S$-factor $S(T'_{Ad})$ is typically taken as an expansion in some powers of the energy, whose
expansion parameters are adjusted to reproduce the available experimental information concerning
the cross section.

\section{Direct capture mechanism}

Let us consider the case where the three-body resonance is formed directly, without population of any intermediate two-body state.  We first developed the general formalism, which afterwards is employed with practical parametrizations.

\subsection{Computational scheme}

The recombination process under consideration takes place in two-steps:
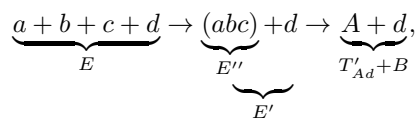
\begin{gather*}
\underbrace {a+b+c+d}_E \rightarrow \underbrace{(abc)}_{E''} + d \rightarrow \underbrace{A+d}_{T'_{Ad}+B}, 
 \\[-\normalbaselineskip] \hspace*{1.3cm}
\underbrace{\kern8mm}_{E'}
\end{gather*}
where the first step is a three-body process where the resonance $(abc)$ is populated,
and the second one is a two-body process where the resonance+$d$ system decays
into $A$+$d$.

Let us denote here by $E$ the total four-body energy, by $E'$ the relative energy between particle $d$ and
the center of mass of the $(abc)$-system, and by $E''$ the three-body energy of the 
$(abc)$-system. It is then clear that
\begin{equation}
E=E'+E''=T'_{Ad}+B=T'_{Ad}-|B|,
\end{equation}
where, as before, $B$ is the binding energy of the three-body system $A$, 
and $T'_{Ad}$ is the final relative kinetic energy between $A$ and $d$.

The reaction rate corresponding to the final two-body step can be found for instance
in Eq.(32) of ref.~\cite{fow67}:
\begin{eqnarray}
\langle R_{abc,d}(E'')\rangle=
\sqrt{\frac{8}{\pi}} \frac{1}{\mu_{abc,d}^{1/2}} \frac{1}{(KT)^{3/2}}  \nonumber && \\ 
\int_0^\infty E' \sigma_{abc,d}(E',E'') e^{-\frac{E'}{KT}} dE',  &&
\label{2bdrr}
\end{eqnarray}
where $K$ again is the Boltzmann constant, $T$ is the temperature, and $\sigma_{abc,d}(E',E'')$
is the two-body cross section for the process $(abc)+d\rightarrow A+d$. 

In order to get the total reaction rate one has to multiply Eq.(\ref{2bdrr}) by the formation probability
 of the three-body $(abc)$-resonance, and average with the Maxwell-Boltzmann
distribution for three particles:
\begin{equation}
B_3(E'',T)=\frac{1}{2}\frac{E''^2}{(KT)^3} e^{-\frac{E''}{KT}}.
\label{mb}
\end{equation} 

The population probability for the three-body resonance is given by the product of the resonance lifetime,
$\tau_{abc} = \hbar/\Gamma_{abc}$, and the rate of formation of the resonance.
This rate is given by the flux of particles times the three-body cross section, $\sigma_{abc}(E'')$.
As shown in  \cite{gar14}, the flux of particles for three-body collisions is given by:
\begin{equation}
\mbox{flux}= \frac{\hbar \kappa''}{m}\left(\frac{m}{\mu_{a,b}}\right)^{3/2} \left(\frac{m}{\mu_{ab,c}}\right)^{3/2},
\end{equation}
where $\kappa''^2=2mE''/\hbar^2$, and $m$ is the usual normalization mass introduced  to construct
the Jacobi coordinates commonly used to describe three-body systems. It is clear that the
flux written above is ill-defined, since it depends on the arbitrary choice of the
normalization mass. However, as we shall show later, the dependence on $m$ disappears in the
final expression of the reaction rate. Therefore, we get for the total reaction rate:
\begin{eqnarray}
&&
\langle R_{abcd} \rangle=\left(\frac{m}{\mu_{a,b}}\right)^{3/2} \left(\frac{m}{\mu_{ab,c}}\right)^{3/2}  
\nonumber \\ && 
\int dE'' B_3(E'',T) \frac{\hbar}{\Gamma_{abc}}
\frac{\hbar \kappa''}{m}
\sigma_{abc}(E'') \langle R_{abc,d}(E'')\rangle ,
\end{eqnarray}
which by use of Eqs.(\ref{2bdrr}) and (\ref{mb}) leads to: 
\begin{eqnarray}
&&
\langle R_{abcd} \rangle=\frac{2\hbar}{\sqrt{\pi}} \frac{m^{5/2}}{(\mu_{a,b} \mu_{ab,c})^{3/2} \mu_{abc,d}^{1/2}} \frac{1}{(KT)^{9/2}} 
  \label{dir1} \\ &&  \!\!
\int dE'' E''^{5/2} \frac{\sigma_{abc}(E'')}{\Gamma_{abc}} e^{-\frac{E''}{KT}}  \!\!\!
\int dE' E' \sigma_{abc,d}(E',E'') e^{-\frac{E'}{KT}}. \nonumber
\end{eqnarray}

Finally, we can now make use of the detailed balance principle that relates $\sigma_{abc,d}$ with the cross section
$\sigma_{Ad}$ for the inverse process, which is:
\begin{equation}
\sigma_{abc,d}=\frac{g_A g_d}{g_{abc} g_d} \frac{\mu_{A,d}}{\mu_{abc,d}} \frac{T'_{Ad}}{E'} \sigma_{Ad},
\label{balan}
\end{equation}
where $g_i=(2J_i+1)$ is the multiplicity of system $i$ with angular momentum $J_i$. Therefore, it
is possible to rewrite Eq.(\ref{dir1}) in terms of $\sigma_{Ad}$ (in principle accessible experimentally) 
and get the final expression for the total reaction rate assuming a direct formation of the three-body
resonance:
\begin{eqnarray}
&&
\langle R_{abcd} \rangle=\frac{g_A}{g_{abc}}
\frac{2\hbar}{\sqrt{\pi}} \frac{m^{5/2} \mu_{A,d}}{(\mu_{a,b} \mu_{ab,c} \mu_{abc,d})^{3/2}} \frac{1}{(KT)^{9/2}} 
   \label{dir2} \\ &&  \!\!
\int dE'' E''^{5/2} \frac{\sigma_{abc}(E'')}{\Gamma_{abc}} e^{-\frac{E''}{KT}}  \!\!\!
\int dE' T'_{Ad} \sigma_{Ad}(T'_{Ad}) e^{-\frac{E'}{KT}}. \nonumber
\end{eqnarray}

All along the derivations, we have not considered the possibility of having identical particles involved in the process.
In order to do so, the expression above should be multiplied by $\prod_i \nu_i!$, where $\nu_i$ is the number
of identical particles of type $i$.

\subsection{Three-body resonant cross section}

In the reaction rate (\ref{dir2}), together with $\sigma_{Ad}$, the key quantity is the three-body resonant
cross section $\sigma_{abc}(E'')$, which is given by \cite{gar18}
\begin{equation}
\sigma_{abc}(E'')=\frac{32 \pi^2}{\kappa''^5} \frac{g_{abc}}{g_ag_bg_c} \frac{\Gamma_{abc}^2}{(E''-E_{abc})^2+\Gamma_{abc}^2/4},
\label{bw3b}
\end{equation}
where $E_{abc}$ is the energy of the three-body resonance. Since  $\kappa''=\sqrt{2mE''}/\hbar$, we now see that
inclusion of Eq.(\ref{bw3b}) into Eq.(\ref{dir2}) eliminates the dependence on the arbitrary normalization mass $m$.

In general, the width $\Gamma_{abc}$ will depend on the three-body energy $E''$, typically as \cite{gar05}:
\begin{equation}
\Gamma_{abc}(E'')=\Gamma_{abc}^{(0)} \frac{P(E'')}{P(E_{abc})},
\label{penf}
\end{equation}
where $P(E'')$ is the penetration factor.

For direct decay of the three-body resonance, and for the case of charged particles, as shown in Eqs.(3) and (12) of \cite{gar05}, 
we take:
\begin{equation}
\Gamma_{abc}(E'')=\Gamma_{abc}^{(0)}\frac{1+e^{2b_{abc}/\sqrt{E_{abc}}}}{1+e^{2b_{abc}/\sqrt{E''}}}
\label{wabc}
\end{equation}
with
\begin{small}
\begin{equation}
b_{abc}=\frac{\pi}{2} \sqrt{\frac{2}{\hbar^2(m_a+m_b+m_c)}}
\left( \sum
\left( Z_i Z_j e^2\right)^{2/3} \left( m_i m_j\right)^{1/3}
\right)^{3/2}
\label{bdir}
\end{equation}
\end{small}
where $m_a$, $m_b$, and $m_c$ are the masses of the particles, $Z_i$ is the charge
of particle $i$,  and the sum runs over the three possible pairs of particles.

Finally, let us consider the extreme case, where the width of the three-body resonance is neglected and taken equal to zero. In this case
we then have that:
\begin{equation}
\frac{\sigma_{abc}(E'')}{\Gamma_{abc}} \stackrel{\Gamma_{abc}\rightarrow 0}{\longrightarrow}
\frac{64 \pi^3}{\kappa_{abc}^5} \frac{g_{abc}}{g_ag_bg_c} \delta(E''-E_{abc}),
\label{extr1}
\end{equation}
where $\kappa_{abc}=\sqrt{2mE_{abc}}/\hbar$, which leads to the zero-width limit of the
reaction rate:
\begin{eqnarray}
&&
\langle R_{abcd} \rangle=\frac{g_A}{g_a g_b g_c}
 \frac{4 (2\pi)^{\frac{5}{2}} \hbar^6 \mu_{A,d}}{(\mu_{a,b} \mu_{ab,c} \mu_{abc,d})^{3/2}} \frac{1}{(KT)^{9/2}} 
\nonumber \\  \label{limit1} &&   e^{-\frac{E_{abc}}{KT}}  \!\!\!
\int dE' T'_{Ad} \sigma_{Ad}(T'_{Ad}) e^{-\frac{E'}{KT}},
\end{eqnarray}
which again should be multiplied by $\prod_i \nu_i!$ ($\nu_i$ is the number
of identical particles of type $i$) in order to account for the presence of identical particles.

Note that in this limit we have that $E=E'+E_{abc}$, and therefore Eq.(\ref{limit1}) is formally identical to
Eq.(\ref{4bdrr}), which was presented as the general form of the four-body recombination rate. 
However, as discussed, in Eq.(\ref{4bdrr}) $\sigma_{Ad}$ describes the full four-body process, and contains
the intrinsic kinematics of the four particles all along the process for all four-body energies, including
the $0\leq E \leq E_{abc}$ low-energy range. On the contrary, in the picture described in this
section, the cross section $\sigma_{Ad}$ is an inelastic two-body process where the nucleus
$A$ is excited into the state with energy $E_{abc}$, which implies that in the extreme limit that led 
to Eq.(\ref{limit1}) only four-body energies satisfying $E\geq E_{abc}$ are allowed. A description
of the reaction rate in the $0\leq E \leq E_{abc}$ energy range requires use of Eq.(\ref{dir2}), being
the resonant cross section $\sigma_{abc}(E'')$ the one dictating the low-energy behaviour of the reaction rate. 

\section{Sequential capture mechanism}

The second possibility to be considered is the case of a sequential mechanism, where, prior to the formation of a three-body resonance,  an intermediate two-body resonance 
is populated.  We first provide the formulation for the combined processes, and afterwards we describe the individual parametrizations.

\subsection{The combined sequence of reactions}

Several steps are involved in this reaction mechanism.  The processes are:
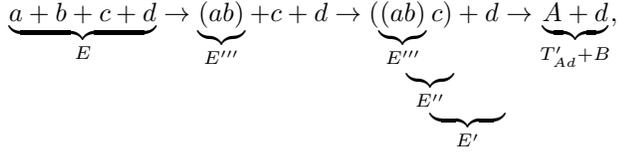
\begin{gather*}
\underbrace {a+b+c+d}_E \rightarrow \underbrace{(ab)}_{E'''}+c+d \rightarrow (\underbrace{(ab)}_{E'''}c) + d \rightarrow \underbrace{A+d}_{T'_{Ad}+B}, 
 \\[-\normalbaselineskip] \hspace*{3.1cm}
\underbrace{\kern5mm}_{E''}
 \\[-\normalbaselineskip] \hspace*{4.1cm}
\underbrace{\kern10mm}_{E'}
\end{gather*}
which are three consecutive two-body processes: The population of the two-body $(ab)$-resonance, the capture of particle $c$ by the two-body $(ab)$-system in order
to form the three-body resonance $(abc)$, and the decay of the resonance into the three-body bound state $A$, whose energy excess is absorbed
by particle $d$. 

In this case the energy $E'''$ is the relative energy between particles $a$  and $b$, and $E''$ is the relative energy between particle
$c$ and the center of mass of the $(ab)$-system (therefore, $E''+E'''$ is the total three-body energy). As before, $E'$ is the relative energy between $d$ and the center of mass of the three-body system,
and $T'_{Ad}$ is the final relative energy between $d$ and the bound system $A$. Therefore the energy conservation reads now:
\begin{equation}
E=E'+E''+E'''=T'_{Ad}+B=T'_{Ad}-|B|.
\end{equation}

The last two-body step is the same as for the direct capture. Therefore, the reaction rate for this last process is
again given by Eq.(\ref{2bdrr}), although now $\sigma_{abc,d}$ depends on the three energies $E'$, $E''$, and $E'''$. 
Again, this rate has to be multiplied by the probability for the formation of the three-body
resonance. The difference is that this probability is now given by  the probability of formation of the two-body $(ab)$-resonance, 
times the probability of the subsequent capture of particle $c$ in order to form the three-body resonance. Each of these two
probabilities is again the product of the resonance lifetime $\tau=\hbar/\Gamma$, the flux of particles $\hbar k/\mu$, and
the corresponding cross section. Each of them has also to be averaged by means of the two-body Maxwell-Boltzmann distribution: 
\begin{equation}
B_2(E,T)=\frac{2}{\sqrt{\pi}}\frac{E^{1/2}}{(KT)^{3/2}} e^{-\frac{E}{KT}}.
\label{mb2}
\end{equation} 

All this leads to the following expression for the total four-body recombination rate in case of a 
sequential capture mechanism:
\begin{eqnarray}
&&
\langle R_{abcd} \rangle=
\int dE''' B_2(E''',T) \frac{\hbar}{\Gamma_{ab}} \frac{\hbar k'''}{\mu_{a,b}} \sigma_{ab}(E''')  
\nonumber \\ && 
\int dE'' B_2(E'',T) \frac{\hbar}{\Gamma_{ab,c}} \frac{\hbar k''}{\mu_{ab,c}} \sigma_{ab,c}(E'',E''')  
\nonumber \\ && 
\langle R_{abc,d}(E'')\rangle ,
\end{eqnarray}
where $k'''=\sqrt{2\mu_{a,b}E'''}/\hbar$ and $k''=\sqrt{2\mu_{ab,c}E''}/\hbar$. Using
now Eqs.(\ref{2bdrr}) and (\ref{mb2}) we then get: 
\begin{eqnarray}
&&
\langle R_{abcd} \rangle=\left(\frac{8}{\pi}\right)^{3/2} \frac{\hbar^2}{(\mu_{a,b} \mu_{ab,c} \mu_{abc,d})^{1/2} } \frac{1}{(KT)^{9/2}} 
  \nonumber \\  && 
\int dE''' E''' \frac{\sigma_{ab}(E''')}{\Gamma_{ab}} e^{-\frac{E'''}{KT}}  
\int dE'' E'' \frac{\sigma_{ab,c}(E'',E''')}{\Gamma_{ab,c}} e^{-\frac{E''}{KT}}
 \nonumber  \\   && 
\int dE' E' \sigma_{abc,d}(E',E'',E''') e^{-\frac{E'}{KT}}. 
\end{eqnarray}

Finally, using again Eq.(\ref{balan}), we get the following expression for the reaction rate of sequential capture in terms of the inelastic cross section $\sigma_{Ad}$:
\begin{eqnarray}
&&
\langle R_{abcd} \rangle=\frac{g_A}{g_{abc}} \left(\frac{8}{\pi}\right)^{3/2} \frac{\hbar^2 \mu_{Ad}}{(\mu_{a,b} \mu_{ab,c})^{1/2} \mu_{abc,d}^{3/2}} \frac{1}{(KT)^{9/2}} 
  \nonumber \\  \label{seq1} && 
\int dE''' E''' \frac{\sigma_{ab}(E''')}{\Gamma_{ab}} e^{-\frac{E'''}{KT}}  
\int dE'' E'' \frac{\sigma_{ab,c}(E'',E''')}{\Gamma_{ab,c}} e^{-\frac{E''}{KT}}
 \nonumber  \\   && 
\int dE' T'_{Ad} \sigma_{Ad}(T'_{Ad}) e^{-\frac{E'}{KT}}. 
\end{eqnarray}

As before, an additional factor $\prod_i \nu_i!$ should be included to account for the presence 
of identical particles.

\subsection{Two-body resonant cross sections}

The cross sections, $\sigma_{ab}(E''')$ and $\sigma_{ab,c}(E'',E''')$, entering in Eq.(\ref{seq1}) are just two-body
cross sections, which for the case of resonance population can be written as:
\begin{equation}
\sigma_{ab}(E''')=\frac{g_{ab}}{g_a g_b} \frac{\pi}{k'''^2} \frac{\Gamma_{ab}^2}{(E'''-E_{ab})^2+\Gamma_{ab}^2/4}
\label{bw2ba}
\end{equation}
and 
\begin{equation}
\sigma_{ab,c}(E'',E''')=\frac{g_{abc}}{g_{ab} g_c} \frac{\pi}{k''^2} \frac{\Gamma_{ab,c}^2}{(E''+E'''-E_{abc})^2+\Gamma_{ab,c}^2/4},
\label{bw2bb}
\end{equation}
respectively, where $E_{ab}$ and $E_{abc}$ are the two-  and three-body resonance energies, whose widths are 
denoted by $\Gamma_{ab}$ and $\Gamma_{ab,c}$, respectively.

Again, the widths are in general energy dependent, a dependence contained in the penetration
factor, as indicated in Eq.(\ref{penf}).  Also, Eq.(\ref{wabc}) is still valid, and we can write down the
energy dependence of the widths as:
\begin{equation}
\Gamma_{ab}(E''')=\Gamma_{ab}^{(0)}\frac{1+e^{2b_{ab}/\sqrt{E_{ab}}}}{1+e^{2b_{ab}/\sqrt{E'''}}}
\label{gab}
\end{equation}
and 
\begin{equation}
\Gamma_{ab,c}(E'')=\Gamma_{ab,c}^{(0)}\frac{1+e^{ 2b_{ab,c}/\sqrt{E_{abc}-E_{ab}} }}{1+e^{2b_{ab,c}/\sqrt{E''}}},
\label{gabc}
\end{equation}
where, for the case of a sequential mechanism, $b_{ab}$ and $b_{ab,c}$ are given by Eqs.(10) and
(11) of \cite{gar05}:
\begin{eqnarray}
b_{ab,c}&=&\frac{\pi}{2}(Z_a+Z_b)Z_c e^2 \sqrt{\frac{2\mu_{ab,c}}{\hbar^2}}
\label{babc}
 \\
b_{ab}&=&\frac{\pi}{2} Z_a Z_b e^2 \sqrt{\frac{2\mu_{ab}}{\hbar^2}},
\label{bab} 
\end{eqnarray}
and where $Z_a$, $Z_b$, and $Z_c$ are the charges of particles $a$, $b$, and $c$, respectively,
and $e$ is the electron charge. 

If particle $c$ is not charged, the penetration factor takes the form given by
Eq.(20) of \cite{gar05}, which gives:
\begin{equation}
\Gamma_{ab,c}(E'')=\Gamma_{ab,c}^{(0)}\left( \frac{E''}{E_{abc}-E_{ab}}\right)^{\ell_{ab,c}+1/2}
\label{gabc1}
\end{equation}
where $\ell_{ab,c}$ is the relative orbital angular momentum between particle $c$
and the center of mass of the $(ab)$-system.

As in the direct case, we consider also the extreme situation where the width of the $(ab)$-resonance is taken equal to 0. This implies that:
\begin{equation}
\frac{\sigma_{ab}(E''')}{\Gamma_{ab}} \stackrel{\Gamma_{ab}\rightarrow 0}{\longrightarrow}
\frac{2\pi^2}{k_{ab}^2} \frac{g_{ab}}{g_ag_b} \delta(E'''-E_{ab}),
\label{extr2}
\end{equation}
where $k_{ab}^2=2\mu_{a,b}E_{ab}/\hbar^2$, and which leads to the following expression
for the zero-width limit of the reaction rate:
\begin{eqnarray}
&&
\langle R_{abcd} \rangle=\frac{g_A}{g_{abc}} \frac{g_{ab}}{g_a g_b} \left(\frac{8}{\pi}\right)^{3/2} \frac{\hbar^4 \pi^2 \mu_{Ad}}{\mu_{a,b}^{3/2} \mu_{ab,c}^{1/2} \mu_{abc,d}^{3/2}} \frac{e^{-\frac{E_{ab}}{KT}} }{(KT)^{9/2}} 
 \label{seq2}  \\ && 
\int dE'' E'' \frac{\sigma_{ab,c}(E'',E''')}{\Gamma_{ab,c}} e^{-\frac{E''}{KT}}
\int dE' T'_{Ad} \sigma_{Ad}(T'_{Ad}) e^{-\frac{E'}{KT}}.  \nonumber
\end{eqnarray}

If, on top of zero width for the $(ab)$-resonance, we also assume zero width for the $(abc)$-resonance, 
we can again make use of Eq.(\ref{extr2}), but applied to the $(ab,c)$ system:
\begin{equation}
\frac{\sigma_{ab,c}(E'')}{\Gamma_{ab,c}} \stackrel{\Gamma_{ab}\rightarrow 0}{\longrightarrow}
\frac{2\pi^2}{k_{ab,c}^2} \frac{g_{abc}}{g_{ab}g_c} \delta(E''-(E_{abc}-E_{ab}) ),
\label{extr3}
\end{equation}
where $k_{ab,c}^2=2\mu_{a,b}(E_{abc}-E_{ab})/\hbar^2$.
After insertion in Eq.(\ref{seq2}) we recover the result in 
Eq.(\ref{limit1}), obtained as the extreme limit of the direct capture case.

\section{Theory remarks}

To make easier a general view of the derivations made so far, we consider it useful to collect the key equations
used in the actual calculations for both direct and sequential capture mechanisms. We also discuss in this 
section how the production rates are computed and how they compare with those applying 
to a radiative capture process.

\subsection{Key equations} 

For the direct capture mechanism we summarize the ingredients: 

\begin{itemize}

\item The general expression for the four-body recombination rate is given by Eq.(\ref{dir2}). 

\item The extreme case of zero width for the three-body resonance 
is given by Eq.(\ref{limit1}).

\item The cross sections contained in the expressions for the reaction rate are given by Eqs.(\ref{sad}) or 
(\ref{sad2}), depending on the charges of $A$ and $d$, and Eq.(\ref{bw3b}).

\item The width entering in Eq.(\ref{bw3b}) is given by  Eq.(\ref{wabc}).
\end{itemize}

For the sequential capture mechanism we summarize the ingredients: 

\begin{itemize}
\item The general expression for the four-body recombination rate is given by Eq.(\ref{seq1}). 

\item The extreme case of zero width for the two-body resonance, $(ab)$, is given by Eq.(\ref{seq2}).

\item Zero width for both the $(ab)$ and the $(ab,c)$-systems gives the same limit as in the direct case, Eq.(\ref{limit1}).

\item The cross sections contained in the expressions for the reaction rate are given by Eqs.(\ref{sad})
or (\ref{sad2}), depending on the charges of $A$ and $d$, Eqs.(\ref{bw2ba}) and (\ref{bw2bb}).

\item The widths entering in Eqs.(\ref{bw2ba}) and (\ref{bw2bb}) are given by  Eqs.(\ref{gab}) and (\ref{gabc}), respectively .

\item If $c$ is not charged, the width in Eq.(\ref{bw2bb}) is given by Eq.(\ref{gabc1}).
\end{itemize}

\subsection{Production rates and radiative capture}
\label{s53}

When comparing the rate corresponding to different reactions, for instance the production of some
nucleus via four-body recombination or three-body radiative capture, this can not in general be done
by simple comparison of the reaction rates. In fact, for a four-body and a three-body process the
corresponding reaction rates have different units, making a direct comparison meaningless. The
comparison has to be done in terms of the production rates, which give the number of reaction
products created per unit time and unit volume.

The production rates are obtained from the reaction rates  after multiplication by the number
density of each of the particles involved in the reaction. 
This density is given by $n_i=\rho N_A X_i/A_i$, where
$\rho$ is the mass density, $N_A$ is the Avogadro number, $X_i$ is the mass abundance of particle $i$, and
$A_i$ is the atomic number of particle $i$.

The ratio between the production rates corresponding to the four-body recombination, $ P_{\mbox{\scriptsize rec}}$,
and the one corresponding to a radiative capture process, $P_{\mbox{\scriptsize rad}}$, leading both to the same
final product, can be then easily computed. In fact,
it is obvious that this ratio, $P_{\mbox{\scriptsize rec}}/P_{\mbox{\scriptsize rad}}$, grows linearly with the number density
of the spectator particle in the four-body reaction. In other words, it grows linearly with the mass density in 
the environment where the reactions take place. Therefore, it is clear that for a sufficiently high value of the
density, the four-body recombination process will dominate over the radiative capture.

In Ref.~\cite{gar11} an analysis similar to the one shown in Sections III and IV was made for the radiative capture process $a+b+c\rightarrow A +\gamma$.
Analogous expressions were obtained for the cases of direct and sequential capture mechanism (Eqs.(3) and (10) in \cite{gar11}).
In particular, it is worth noticing that Eq.(15) of Ref.~\cite{gar11} gives a simple analytical expression for the 
radiative capture process in the limit of zero width for the two- and three-body resonances.
This expression is therefore the equivalent to Eq.(\ref{limit1}) for the four-body recombination reaction in the same limit.
Therefore, use of the reaction rates in Eq.(\ref{limit1}) and in Eq.(15) of Ref.~\cite{gar11} permits us to obtain a simple
expression for $P_{\mbox{\scriptsize rec}}/P_{\mbox{\scriptsize rad}}$ in the extreme case of making the widths of the
resonances equal to zero. This ratio takes the form:
\begin{eqnarray}
\lefteqn{
\frac{P_{\mbox{\scriptsize rec}}}{P_{\mbox{\scriptsize rad}}} \stackrel{\Gamma_{abc}\rightarrow 0}{\longrightarrow}
} \label{austin}  \\ &&
n_d \frac{g_A}{\mu_{Ad}^{1/2}} 
\sqrt{\frac{8}{\pi}} \frac{\hbar}{\Gamma_\gamma(E_{abc})} \frac{1}{(KT)^{3/2}} \int_0^\infty dE' T'_{Ad}\sigma_{Ad}(T'_{Ad}) e^{-\frac{E'}{KT}},
\nonumber
\end{eqnarray}
where $n_d$ is the number density of the spectator particle $d$, $\tau_\gamma=\hbar/\Gamma_\gamma(E_{abc})$ is the lifetime
for gamma decay of the $(abc)$-resonance, and $T'_{Ad}=E'-(B-E_{abc})$.

The expression (\ref{austin}) is the one actually used in Refs.~\cite{tru69,mor70,dav71,bea17} for the production rate ratio between the particle 
induced de-excitation process and the radiative capture process. It is then clear that in those works the effect of the
resonance widths and the reaction mechanism are neglected, in such a way that the only key ingredient entering is the 
$\sigma_{Ad}$ cross section. We can then conclude that, provided the same $\sigma_{Ad}$ cross sections are used, the results in \cite{tru69,mor70,dav71,bea17}
will be the same as obtained in this work in the limit of zero resonance widths. As we will see, the effect of the resonance
widths  will be visible only in the low temperature region, which implies that for high temperatures, where Eq.(\ref{austin}) is a good
approximation, no relevant differences between our results and the ones in  Refs.~\cite{tru69,mor70,dav71,bea17} are expected.

Finally, let us close this section mentioning that in Eq.(\ref{austin}) it is assumed that the four-body and the three-body
processes involve the same number of identical particles (for instance the $\alpha+\alpha+\alpha+n\rightarrow \mbox{$^{12}$C}+n$
and $\alpha+\alpha+\alpha \rightarrow \mbox{$^{12}$C}+\gamma$ processes). However, other situations are possible, like  having four
identical particles in the four-body recombination process (for instance the $\alpha+\alpha+\alpha+\alpha\rightarrow \mbox{$^{12}$C}+\alpha$
process), in which case the ratio in Eq.(\ref{austin}) should be multiplied by 4. In general, Eq.(\ref{austin}) has to be multiplied
by the factor $\prod_i \nu_i!$, corresponding to the four-body particle induced de-excitation process, and divided by the same 
factor, $\prod_i \nu_i!$,
corresponding to the three-body radiative process. As before, for each of the two reactions, $\nu_i$ is the number
of identical particles of type $i$.

\section{The cases of $^9$Be and $^{12}$C}

In this section we apply the results previously described to four-body recombination processes leading to the formation of 
$^9$Be ($\alpha+\alpha+n$) and $^{12}$C ($\alpha+\alpha+\alpha$). The cases where a neutron, a proton, or an $\alpha$ particle 
act as spectator will be considered.

\subsection{Cross sections}
\label{seccs}

The crucial ingredients for the calculation of the reaction rates are the cross sections contained in the expressions (\ref{dir2}) 
and (\ref{seq1}) corresponding to the direct and sequential mechanisms, respectively.

As mentioned, two different types of cross sections appear. The first type corresponds to population of a two- or a three-body resonance,
which are given by Eq.(\ref{bw3b}) for direct capture, and Eqs.(\ref{bw2ba}) and  (\ref{bw2bb}) for sequential capture. These cross sections
are fully specified by the corresponding resonance energies and widths.

For the two cases considered in this work, $^9$Be and $^{12}$C, we shall include in the calculation the contribution from the two
three-body resonances contributing the most to reaction rates. As in the case of radiative capture \cite{gar11}, the effect of the second 
resonance is expected to play a role only for high temperatures. In this temperature region there could also be some small effects from 
even higher resonances, which could interfere through the resonance tails. This effect is not very relevant for the purpose of this work, 
and therefore it will not be considered here.

\begin{table}
\begin{center}
\begin{tabular}{c|c|cc|cc}
  &  $^8$Be   &  \multicolumn{2}{c}{$^9$Be}  & \multicolumn{2}{|c}{$^{12}$C}  \\ \hline
 $ J^\pi$ &   $0^+$    &  $1/2^+$   &   $5/2^+$    &   $0^+$    &  $2^+$  \\
$E_{\mbox{\scriptsize res}}$ &    0.092   &   0.110   &   1.475   &  0.38  &  1.84 \\
$\Gamma_{\mbox{\scriptsize res}}$  &  $5.6 \times 10^{-6}$ & 0.217  &  0.282  &  $9.3\times 10^{-6}$ & 0.56  \\
\end{tabular}
\end{center}
\caption{Resonance energies, $E_{\mbox{\scriptsize res}}$, and widths, $\Gamma_{\mbox{\scriptsize res}}$, in MeV, for the
two- and three-body resonances included in the calculations of the reaction rates in Eq.(\ref{dir2}), direct capture, and Eq.(\ref{seq1}), sequential
capture, for the formation of $^9$Be and $^{12}$C. }
\label{tab1}
\end{table}

In the case of $^9$Be, we shall consider the contributions to the rate from the 1/2$^+$ and 5/2$^+$ three-body resonances
decaying into the 3/2$^-$ ground state of $^9$Be. The contribution from the $1/2^+$ state is expected to clearly dominate, although, 
as shown in \cite{die14}, the contribution from the 5/2$^+$ resonance could be important at high temperatures.
The energies and widths $(E_{\mbox{\scriptsize res}}, \Gamma_{\mbox{\scriptsize res}})$
employed for the 1/2$^+$ and 5/2$^+$ resonances are, respectively, (0.110, 0.217) MeV and (1.475, 0.282) MeV \cite{til04}. The binding
energy of the $^9$Be ground state is $B=-1.574$ MeV \cite{til04}. 

In case of sequential capture, the process is assumed to take place through
the $0^+$ resonance in $^8$Be, whose energy and width are, respectively, 0.092 MeV and $5.6\times 10^{-6}$ MeV \cite{til04}.
In principle, the sequential formation of $^9$Be could also take place through the lowest, $\frac{3}{2}^-$, resonance in $^5$He ($\alpha+n$). 
However, this resonance is located at 0.798 MeV above threshold \cite{til02}, being then much less energetically accessible than the 
 significantly lower 0$^+$ resonance in $^8$Be.
In fact, even for temperatures of 1 GK the Maxwell-Boltzmann distribution peaks at an energy
even below 0.5 MeV. Furthermore, the large width of the  $\frac{3}{2}^-$ resonance in $^5$He (0.648 MeV, \cite{til02}) makes the
capture of the second alpha particle before the resonance decay very unlikely, and, at the same time, smears out the resonance contribution
to the reaction rate into the continuum background. For these reasons the sequential capture through $^5$He intermediate states
will not be considered here.

For $^{12}$C, the contributions to the rate from the 0$^+$ and 2$^+$ three-body resonances
decaying into the 0$^+$ ground state  will be considered. The energy and width 
employed for the Hoyle 0$^+$ state are  (0.38, $9.3\times 10^{-6}$) MeV \cite{kel16}. For the lowest $2^+$ resonance
we take (1.84, 0.56) MeV, as suggested in \cite{ang99}. The required binding energy of the $^{12}$C ground state is 
$B=-7.275$ MeV \cite{kel16}. In case of sequential capture, 
the process is again taking place through the $0^+$ resonance in $^8$Be. The energies and widths of the resonances included in the calculation are collected in Table~\ref{tab1}.

The second type of cross sections are the inelastic $\sigma_{Ad}$ cross sections.  They will be assumed to
take the form given in Eqs.(\ref{sad}) and ($\ref{sad2}$) for the cases where particle $d$ is charged or neutral, respectively. 
Furthermore, we know that the cross section must be equal to zero for collision energies smaller than $|B|$, where $B$ is the binding
energy of the $A$ system. To guarantee this condition, we multiply the cross sections  (\ref{sad}) and ($\ref{sad2}$) by the
shape function:
\begin{equation}
f(E)=\frac{e^{(E-|B|)/a}-1}{e^{(E-|B|)/a}+1},
\label{difu}
\end{equation}
which is 0 for $E=|B|$ and 1 for $E \gg  |B|$. The diffuseness parameter $a$ determines how fast the transition from 0 to 1 takes place.

\begin{table}
\begin{center}
\begin{tabular}{c|cc|cc|cc|}
  & \multicolumn{2}{c}{proton}  & \multicolumn{2}{|c}{neutron}  &  \multicolumn{2}{|c|}{alpha}  \\ 
   &  $^9$Be & $^{12}$C  &  $^9$Be & $^{12}$C &  $^9$Be & $^{12}$C  \\ \hline
$S_0$  &  10   &   10  &  0.1     &  0.1  &   10$^3$ &   10$^3$  \\
$\gamma$  &   0  &  $-0.032$   &  0    & $-0.040$   &  $-0.045$   &    $-0.010$\\
$a$  & 0.9     &  1.5    &  0.9    & 1.5    &   0.1  &  8.0  \\
\end{tabular}
\end{center}
\caption{Parameters used for the inelastic cross sections $\sigma_{Ad}$ for the cases of incident proton, neutron, and alpha  on $^9$Be and $^{12}$C.
$S_0$ is given in units of MeV·barn for incident protons and alphas, and in units of MeV$^{1/2}$·barn for the case of incident neutron.
The parameters $\gamma$ and $a$ are given in MeV$^{-1}$ and fm, respectively.}
\label{tab2}
\end{table}

The key point here is how to determine the $S$-factors, which should be estimated for the cases of incident proton, neutron, or alpha
particle on both, $^9$Be and $^{12}$C. In this work we shall take $S(E)=S_0(1+\gamma E)$, where
$S_0$ and $\gamma$ are constants to be extracted from the available experimental information.

\begin{figure}[t!]
\psfig{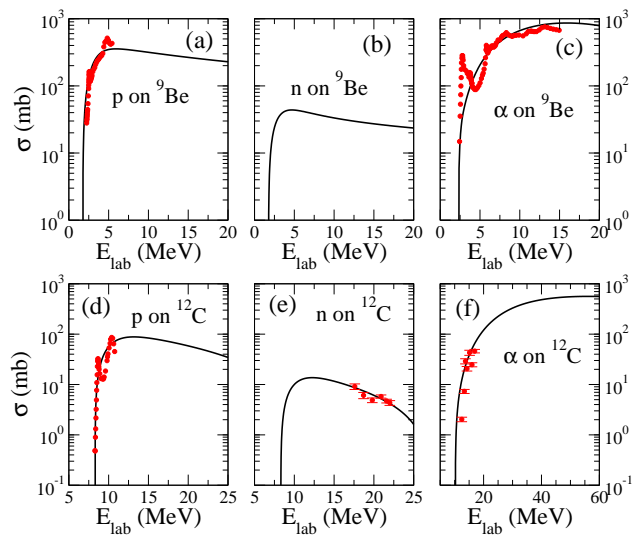}
\vspace*{-1mm}
\caption{Cross sections, in mb, as a function of the lab energy of the incident particle, for proton, neutron, and alpha on $^{9}$Be, panels (a), (b),
and (c), respectively, and for proton, neutron, and alpha on $^{12}$C, panels (d), (e), and (f), respectively. The experimental data (solid circles)
are from  \cite{gib59}  and \cite{gib65} in panels (a) and (c), and from Refs.\cite{dav71}, \cite{ols89},  and \cite{mor70},
in panels (d), (e), and (f), respectively. }
\label{fig1}
\end{figure}

In Ref.~\cite{ang99}, the $S$-factor for quite a few reactions are plotted. Among them the one for the reaction $^9$Be($p,n$)$^9$B.
The $S$-factor corresponding to the proton-$^9$Be collision grows fast from zero and it soon reaches a rather constant value of about
10 MeV·barn. This value seems not to depend much on the charge of the target nucleus, as seen for instance for the $^{11}$B($p,n$)$^{11}$C
and $^{13}$C($p,n$)$^{13}$N reactions, where a similar constant value of the $S$-factor is seen. Therefore, for the case of incident proton we take
$S_0= 10$ MeV·barn for both, $^9$Be and $^{12}$C. This value of $S_0$ together with a diffuseness parameter
$a=0.9$ fm in Eq.(\ref{difu}), are enough to reproduce reasonably well the experimental values of the cross section given in \cite{gib59} for the case of $^9$Be
target. The computed cross section (solid line) and the experimental data (solid circles) are shown in Fig.~\ref{fig1}a. For the case of 
$^{12}$C, the comparison with the experimental data in \cite{dav71}, Fig.~\ref{fig1}d, suggests a bigger value of the diffuseness
parameter, $a=1.5$ fm, and a small correction of the $S$-factor by taking $\gamma=-0.032$ MeV$^{-1}$. 
The introduction of $\gamma$ gives rise to  a fall-off of the cross section for large energies similar to the one shown in \cite{bea17}.

For the case of incident neutron the value of $S_0$ is given in Ref.~\cite{fow67} for several different reactions. However, its value oscillates
quite a lot from one reaction to another. For instance for the  reactions $^{7}$Be($n,p$)$^{7}$Li and  $^{14}$N($n,p$)$^{14}$C the difference in the
given $S_0$ is of almost five orders of magnitude. As a consequence, it is not easy to extract a more or less reliable value to be used
in our calculations. Therefore, for each of the two cases, $^9$Be and $^{12}$C, we have opted for using the same diffuseness parameter $a$ as in the proton
case, and fine tune $S_0$ and $\gamma$ such that the experimental values given in 
Ref.~{\cite{ols89} for the case of $^{12}$C are reasonably well reproduced. This is done by taking $S_0=0.1$ MeV$^{1/2}$·barn and $\gamma=-0.04$ MeV$^{-1}$,
as shown in Fig.~\ref{fig1}e. Thus, for the case of $^9$Be we have also used $S_0=0.1$ MeV$^{1/2}$·barn, whereas $\gamma$ is taken equal to zero 
as in the proton case. For completeness we show in Fig.~\ref{fig1}b the corresponding cross section.

Finally, in Ref.~\cite{ang99} the $S$-factor for the  $^9$Be($\alpha ,n$)$^{12}$C reaction is also shown. In this case the $S$-factor is not as constant
as a function of the energy as in the proton case, but for an energy  in the vicinity of 1.5 MeV (needed to excite the $^9$Be nucleus), it takes
a value that oscillates around $10^3$ MeV·barn. A similar value is observed for the $^{13}$C($\alpha ,n$)$^{16}$O and the 
$^{14}$N($\alpha ,n$)$^{17}$F reactions. Therefore, for the alpha particle  we take $S_0= 10^3$ MeV·barn 
for both $^9$Be and $^{12}$C. The main difference between these two cases is that in order to reproduce the experimental cross
sections (Ref.~\cite{gib65} for $^9$Be and Ref.~\cite{mor70} for $^{12}$C) a quite small value of $a$ is needed for $^9$Be ($a=0.1$ fm) producing a
very sharp increase of the cross section at small energies. However,  for $^{12}$C the increase of the cross section is much smoother, and
actually a rather large value of $a$ is needed ($a=8$ fm). These values of the diffuseness parameter, together with $\gamma=-0.045$ MeV$^{-1}$ for
$^9$Be and  $\gamma=-0.010$ MeV$^{-1}$ for $^{12}$C, give rise to the cross sections shown by the solid curves in Figs.~\ref{fig1}c and Fig.~\ref{fig1}f,
respectively, where the experimental data are given by the solid circles. All the parameters used for the different inelastic cross sections discussed above are collected in Table~\ref{tab2}.

It is important to note that the experimental data shown in Fig.~\ref{fig1} obviously correspond to collisions of the alpha, proton, or
neutron projectile on the target in its ground state. The uncertainties introduced by using the same cross sections for the case
of collisions on an excited state of the target (like for instance the bound excited 2$^+$ state in $^{12}$C) are big, and could lead to 
very unrealistic results. For this reason we have considered here decays of the three-alpha resonances into the ground state of
$^{12}$C only.

\subsection{Results for $^9$Be}
 
Let us now make use of the cross sections described in the previous section and consider the four-body recombination
processes  $\alpha+\alpha+n+\alpha \rightarrow \mbox{$^9$Be}+\alpha$, $\alpha+\alpha+n+n \rightarrow \mbox{$^9$Be}+n$,
and $\alpha+\alpha+n+p \rightarrow \mbox{$^9$Be}+p$, where either an alpha particle, or a neutron, or a proton, respectively, act
as spectators inducing the formation of $^9$Be.

\begin{figure}[t!]
\psfig{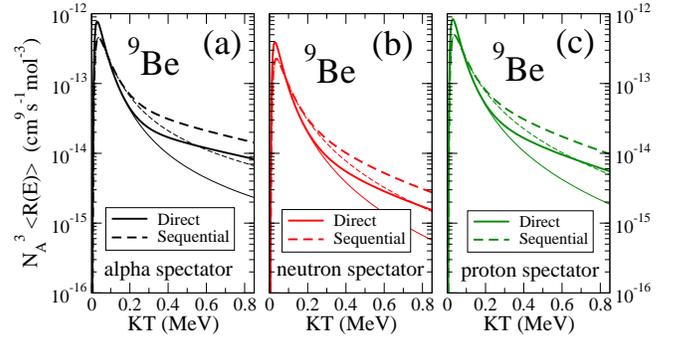}
\caption{Four-body recombination rates for (a) the $\alpha+\alpha+n+\alpha \rightarrow \mbox{$^9$Be}+\alpha$,
(b) the $\alpha+\alpha+n+n \rightarrow \mbox{$^9$Be}+n$, and (c) the $\alpha+\alpha+n+p \rightarrow \mbox{$^9$Be}+p$
reactions as a function of $KT$ (in MeV).
The solid and dashed curves show the cases of direct and sequential capture, respectively. The corresponding thin curves give the reaction
rate including the contribution from the 1/2$^+$ resonance only. $N_A$ is the Avogadro's number.}
\label{fig2}
\end{figure}

In Fig.~\ref{fig2} the solid curves give, for each of the three cases, panels (a), (b), and (c), respectively, the four-body recombination rate obtained assuming a direct capture
mechanism, as considered when deriving  Eq.(\ref{dir2}). In the same way, the dashed curves show the same rates but assuming a sequential mechanism, i.e., using
Eq.(\ref{seq1}). Whereas the thick curves contain the contribution form both, the 1/2$^+$ and the $5/2^+$ resonances in $^9$Be, the thin curves
contain the contribution from the 1/2$^+$ resonance only.

As we can see, in all the three cases, alpha, neutron, and proton spectator,  
the direct mechanism dominates for very small energies, but for values of $KT$ larger than about 0.1 MeV ($T\approx 1 GK$) the sequential 
mechanism always dominates. We also observe that the cases of alpha and proton as spectator produce similar reaction rates, which
dominate over the case of having a neutron as spectator particle. Finally, we also see that the effect of the 5/2$^+$ resonance contribution
is clearly visible in all the cases and capture mechanisms for $KT$ values beyond about 0.2 MeV.

The format in Fig.~\ref{fig2} has been chosen to make easier the comparison with the right part of Fig.~3 in Ref.~\cite{die14}. As we can see the 
result obtained here is consistent with the one shown in \cite{die14} for the 1/2$^+$ resonance, which is by far the dominating contribution.
This is like this in  both cases, when the alpha particle is the spectator (thin-dashed curve in \cite{die14}), and when the neutron is the spectator 
(thick-dashed curve in \cite{die14}). The case of proton spectator was not considered in Ref.~\cite{die14}.

\begin{figure}[t!]
\psfig{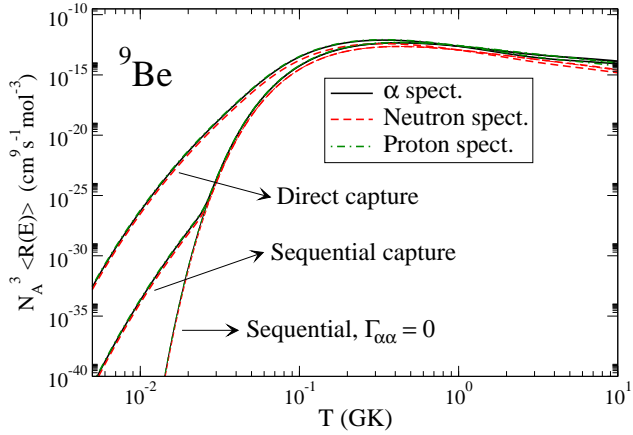}
\vspace*{-1mm}
\caption{Same reaction rates as in Fig.~\ref{fig2} as a function of $T$ and logarithmic scale. The solid, dashed, and dot-dashed curves
show the results obtained with an alpha, a neutron, or a proton as spectator particle. The cases of direct, sequential, and 
extreme sequential ($\Gamma_{\alpha \alpha}=0$, Eq.(\ref{seq2})) capture are shown. $N_A$ is the Avogadro's number.}
\label{fig3}
\end{figure}

In order to make easier the comparison between the different curves at very low temperatures we show in Fig.~\ref{fig3} the
same reaction rates as in Fig.~\ref{fig2} but using a logarithmic scale in the $x$-axis, where the temperature (in GK) is given. In the figure we also
show the extreme cases where the width of the two-body $(\alpha \alpha)$-resonance is taken equal to zero in the case of 
a sequential process. 
We can see that for low temperatures the rate corresponding to a direct mechanism can be  up to about 8
orders of magnitude bigger than the one corresponding to the sequential process. Also, a comparison between the 
extreme sequential ($\Gamma_{\alpha\alpha}=0$) and the sequential cases
permits us to observe that the effect of the width of the $(\alpha \alpha)$-resonance is visible only for 
temperatures smaller than $\approx 0.02$ GK, whereas for higher  temperatures the effect of the two-body resonance width is negligible.  
In any case, the dominance of one capture mechanism versus the other could lead to an important change of the reaction rate at low
temperatures of several orders of magnitude, as also found in Ref.~\cite{gar11} for the case of radiative capture.

\subsection{Results for $^{12}$C}

Let us now consider the processes leading to the formation of $^{12}$C, i.e. the reactions $\alpha+\alpha+\alpha+\alpha \rightarrow \mbox{$^{12}$C}+\alpha$,
$\alpha+\alpha+\alpha+n \rightarrow \mbox{$^{12}$C}+n$, and $\alpha+\alpha+\alpha+p \rightarrow \mbox{$^{12}$C}+p$

\begin{figure}[t!]
\psfig{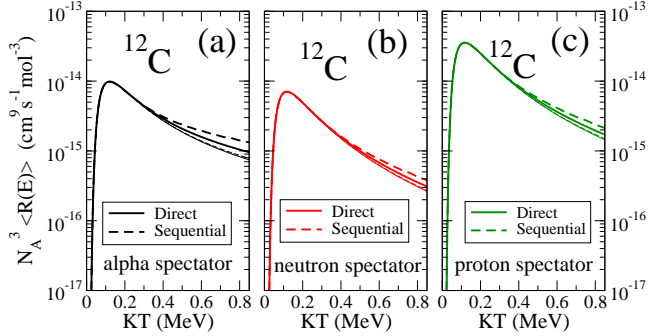}
\caption{Four-body recombination rates for (a) the $\alpha+\alpha+\alpha+\alpha \rightarrow \mbox{$^{12}$C}+\alpha$, (b) the
$\alpha+\alpha+\alpha+n \rightarrow \mbox{$^{12}$C}+n$, and (c) the $\alpha+\alpha+\alpha+p \rightarrow \mbox{$^{12}$C}+p$
reactions as a function of $KT$ (in MeV).
The solid and dashed curves show the cases of direct and sequential capture, respectively. The corresponding thin curves give the reaction
rate including the contribution from the Hoyle 0$^+$ resonance only. $N_A$ is the Avogadro's number.}
\label{fig4}
\end{figure}

The results are shown in Fig.~\ref{fig4}, where again the solid and dashed curves represent, respectively, the reaction rates
obtained assuming a direct or a sequential capture mechanism. The thin curves give the result including only the transition
from the Hoyle 0$^+$ resonance. From the figure we can emphasize the following results. First, we see now a much 
smaller variation compared to the case of $^9$Be between the direct and sequential pictures, although still the sequential
mechanism tends to dominate at high temperatures. Second, the peak of the reaction rates is even more than an order of magnitude smaller than
for $^9$Be. Third, the case of having a proton as spectator produces now a clearly bigger rate than when the spectator is either a 
neutron  or an alpha particle. For the case of $^9$Be the rates obtained  for the proton and alpha spectators were very similar to each other.
And finally, fourth, similarly to what happens with the 5/2$^+$ resonance in $^9$Be, for $^{12}$C the contribution of the $2^+$ resonance
shows up for values of $KT$ beyond 0.4 MeV, although in this case the increase produced in the reaction rate is more modest.

\begin{figure}[t!]
\psfig{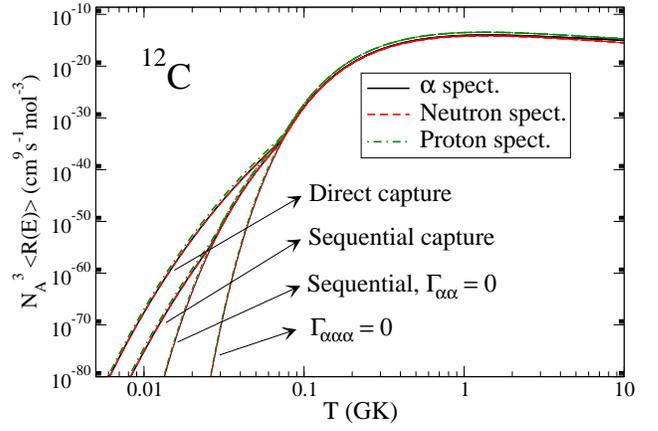}
\vspace*{-1mm}
\caption{Same reaction rates  as in Fig.~\ref{fig4} as a function of $T$ and logarithmic scale. 
The solid, dashed, and dot-dashed curves
show the results obtained with an alpha, a neutron, or a proton as spectator particle.  Together with the cases of direct, sequential, and 
extreme sequential ($\Gamma_{\alpha \alpha}=0$, Eq.(\ref{seq2})) capture mechanisms, we show here as well the extreme case
of zero-width for the three-body resonance ($\Gamma_{\alpha\alpha\alpha}=0$, Eq.(\ref{limit1})).  $N_A$ is the Avogadro's number.}
\label{fig5}
\end{figure}

We again emphasize the differences between the different scenarios at low temperatures by introducing a logarithmic scale
in the temperature axis. This is shown in Fig.~\ref{fig5}, where the meaning of the different curves is as in Fig.~\ref{fig3}, although
in this case we have included as well the limit in Eq.(\ref{limit1}), obtained assuming 
the width of the three-body resonance (the Hoyle state in our case) equal to zero. 
As we can see, at high temperatures (above 0.1 GK) the distinction between the results obtained with the sequential 
and direct mechanisms is hardly seen.
Below this temperature we observe an effect of the three-body resonance width, as indicated by the jump in the reaction
rates corresponding to the $\Gamma_{\alpha\alpha\alpha}=0$ limit and the other scenarios. For temperatures below about 
0.02 GK the effect of the two-body resonance width is seen in the sequential case (jump from the $\Gamma_{\alpha\alpha}=0$ curves
and the ones corresponding to the full sequential capture mechanism).

All in all, again a dominance of the direct capture mechanism  can produce a significant increase of the reaction
rates at very low temperatures, increase that can reach up to seven orders of magnitude compared to the case of dominance
of the sequential capture mechanism.

\subsection{Productions rates}

As discussed in Section~\ref{s53}, the production rates are obtained after multiplication of the reaction rates
by the  density of each of the particles involved in the reaction, i.e.,  $n_i=\rho N_A X_i/A_i$, where
$\rho$ is the mass density, $N_A$ is the Avogadro number, $X_i$ is the mass abundance of particle $i$, and
$A_i$ is the atomic number of particle $i$. This therefore implies that the ratio between the four-body
recombination and the radiative three-body production rates will be proportional to the density $n_d$ of the spectator
particle $d$ in the four-body process. In other words, it will be proportional to the mass density $\rho$ and the mass abundance $X_d$ of the
spectator particle. 

In what follows we shall take $\rho=1$ g/cm$^3$ and $X_d=1$, in such a way that to get the production rate ratio
for given values of $\rho$ and $X_d$ one should simply multiply the computed curves by $\rho X_d$ (with $\rho$ given
in g/cm$^3$).

\begin{figure}[t!]
\psfig{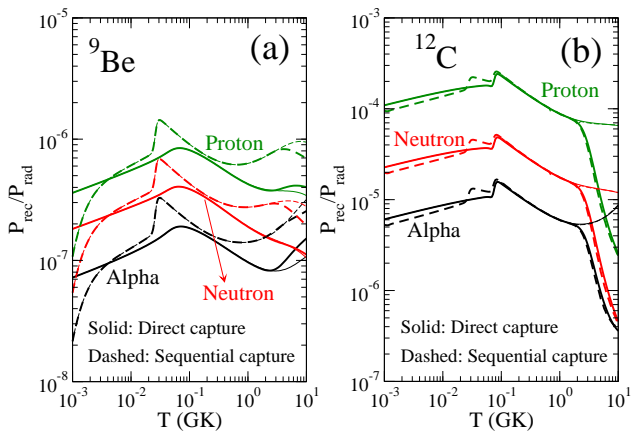}
\caption{Ratio between the four-body recombination production rates giving rise to $^9$Be, panel (a), and
$^{12}$C, panel (b), and the respective radiative capture processes  $\alpha+\alpha+n \rightarrow \mbox{$^9$Be} + \gamma$
and $\alpha+\alpha+\alpha \rightarrow \mbox{$^9$Be} + \gamma$. A mass density $\rho=1$ g/cm$^3$ and spectator
abundance  $X_d=1$ are assumed.
In both panels, the solid and dashed curves correspond to direct and sequential capture mechanisms, respectively.
The cases of an alpha, a neutron, or a proton as spectator in the four body process are shown. The thin curves are obtained 
excluding the contributions from the 5/2$^+$ resonance in the $^9$Be case, and the 2$^+$ resonance in the $^{12}$C case.}
\label{fig6}
\end{figure}

In Fig.~\ref{fig6}a we show, as a function of the temperature, the ratio between the production rates corresponding to the four-body recombination
processes giving rise to $^9$Be, and the production rate corresponding to the radiative capture process
$\alpha+\alpha+n \rightarrow \mbox{$^9$Be} + \gamma$. We take the radiative production rates obtained 
as described in Ref.~\cite{gar11}.  The solid and dashed curves show, respectively, the results assuming a direct or a sequential
capture mechanism. The cases of  alpha, neutron, or proton as spectator particle in the four-body reaction are shown.

As we can see, for the mass density used in the calculation, $\rho=1$ g/cm$^3$, the particle induced recombination
is far of dominating. As anticipated in Ref.~\cite{die14}, densities of about $\rho=10^6$ or $10^7$ g/cm$^3$ are
needed for the four-body recombination to dominate (provided that the abundance $X_d$ of the spectator
particle is close to 1). This is particularly true for high temperatures in case of sequential
capture. The kink observed at about 0.02 GK for the sequential mechanism (dashed curves) is a consequence of the
kink observed at the same temperature in the reaction rates, as seen in Fig.~\ref{fig3}. 
We also show in the
figure the ratio obtained after excluding the contribution from the 5/2$^+$ resonance (thin curves). Only a minor effect
is seen for temperatures close to 1 GK.

In Fig.~\ref{fig6}b the same ratio is shown but for the reactions leading to the formation of $^{12}$C.
The meaning of the curves is as in Fig.~\ref{fig6}a. Comparing with the $^9$Be case we observe that 
in this case the difference between the direct and sequential capture mechanisms is now much smaller.
This fact was already observed in Fig.~\ref{fig4}, where the reaction rates are shown to depend on the
reaction mechanism much less that in the $^9$Be case (Fig.~\ref{fig2}). 

Another relevant feature observed in Fig.~\ref{fig6}b
is that the production rate ratio is now about two orders of magnitude bigger than in panel (a) for all the cases.
This implies that densities about two orders of magnitude
smaller, $10^4$ or $10^5$ g/cm$^3$, could be sufficient to make the four-body process dominate (again assuming
$X_d \approx 1$).  This is especially true at intermediate energies, since for energies approaching 1 GK a rather
pronounced drop of the ratio is found. This is due to the fact that for these temperatures the contribution
from the $2^+$ resonance in $^{12}$C becomes important, and it has a clearly more significant
weight for the radiative capture process than for the particle induced recombination. The consequence is 
the observed fall in the ratio. This can be better understood by looking at the same ratio but excluding the contribution 
of the $2^+$ resonance from the calculation. In this case, thin curves in the figure, the drastic decrease
of the ratio disappears. It is then clear that an unambiguous determination of the energy and width of the $2^+$ resonance
in $^{12}$C is crucial in order to determine the production rate ratio at high temperatures.

As discussed, our results for high temperatures, where the zero-width limit given by Eq.(\ref{austin}) is a good approximation,
should be consistent with the previous results for $^{12}$C shown in Refs.~\cite{mor70,dav71,bea17}, where Eq.(\ref{austin}) is
actually used. From Fig.~\ref{fig6}b we can see that for temperatures around 10 GK and a mass density of $\rho=10^6$ g/cm$^3$,
in the case of neutron or proton as spectator, the production rate for the four-body recombination process is 
within factors of 10 to 100 times the production rate for the radiative capture process, which is consistent with the value of about 
20 to 30 given in \cite{dav71,bea17}. For the case of alpha spectator and the same mass density, the same factor ranges within 1 to 10,
which can also be considered consistent with the results in \cite{mor70} and \cite{bea17}, where they report  values of about 3 and 0.3, 
respectively. The differences between our results and the ones in \cite{mor70,dav71,bea17} can be easily understood as
due to the uncertainties in the $\sigma_{Ad}$ cross sections 
(Fig.~\ref{fig1}), and the ones arising by the role played by the 2$^+$ resonance, as illustrated by the difference between the
thick and thin curves in Fig.~\ref{fig6}b at high temperatures.  At low temperatures, however, where Eq.(\ref{austin}) does not
hold anymore, our results and the ones in 
\cite{mor70,dav71,bea17} are very different. Whereas for $T=10^{-3}$ GK we predict production
rate ratios that for  $\rho=10^6$ g/cm$^3$ range from 10 to 100, the corresponding ratio in \cite{mor70,dav71,bea17}
is essentially zero. This is the consequence of taking into account the combined effect of the resonance width and the 
possibility of a direct capture mechanism.

Finally, let us mention that in Fig.~\ref{fig6}b we observe one kink in the curves at about 0.08 GK for both, the
direct and sequential mechanisms, and an additional one at about 0.02 GK for the sequential case. They reflect the 
corresponding kinks observed in Fig.~\ref{fig5}, at the same temperatures, for each of the capture
mechanisms.

\section{Summary and conclusions}

In this work we have derived analytical expressions for the particle induced recombination reaction rates taking into account
the two possible capture mechanisms, sequential and direct, with and without population of an intermediate two-body state.

The derived reaction rates contain several cross sections that are described through simple phenomenological expressions. In particular,
for those cross sections describing the population of either a two-body or a three-body resonance a Breit-Wigner form is used.
For the cross sections describing the inelastic excitation of the final bound product a standard energy-dependent form
in terms of the $S$-factor is employed.

We apply the theoretical framework to the case of production of $^9$Be and $^{12}$C in an environment containing
neutrons, protons, and alpha particles. The contribution from the 1/2$^+$ and $5/2^+$ resonances are considered for the case of 
$^9$Be. For $^{12}$C, together with the Hoyle state, the contribution from the $2^+$ resonance is also included. When available, the
experimental energy and width of the resonances are used as input for the calculation. Otherwise, like the 2$^+$ resonance
in $^{12}$C, a theoretical estimate is employed. The required inelastic cross sections for alpha, neutron, or proton scattering
 on $^9$Be and $^{12}$C are also constructed to reproduce the available experimental data.

As a general result, we have found that the two possible reaction mechanisms do not produce very different reaction rates,
neither for $^9$Be nor $^{12}$C, at intermediate and high temperatures, although the sequential process is always producing 
a higher rate at high temperatures. However, at very low temperatures, the rate is highly sensitive to both, the width of the
resonances and the capture mechanism. In particular, a direct capture mechanism can produce reaction rates about seven or 
eight orders of magnitude higher than for the sequential mechanism.

Compared to the radiative production rates, we have found that the particle induced recombination can become dominant
for mass densities in the vicinity of 10$^6$ or $10^7$ g/cm$^3$ for the case of $^9$Be. This result is consistent with previous
estimates. For $^{12}$C, the required mass density appears to be about two orders of magnitude smaller, in the vicinity
of 10$^4$ or $10^5$ g/cm$^3$.

Let us finish by emphasizing that the results obtained in this work are based on the reliability of the experimental 
information that permits to construct reliable resonant and non-resonant cross sections. In fact, a very detailed and accurate 
experimental information of the $\sigma_{Ad}$ cross sections, especially at very small energies, would be enough to obtain
accurate values for the reaction rates through Eq.(\ref{4bdrr}). However, since these experimental data are very scarce, and at 
very small energies even very unlikely, we believe the analysis developed in this work can be a relevant tool providing
the input in future astrophysical abundance calculations, where the important effects of nuclear four-body recombination are included.  We would be pleased and also not surprised, 
if significant impact is encountered. However, this obvious perspective requires specialists for reliable implementation, and we therefore leave such application for future work.

\begin{acknowledgements}
This work has been partially supported by
the Spanish Ministry of Science, Innovation and University
MCIU/AEI/FEDER,UE (Spain) under Contract No. PGC2018-093636-B-I00.
\end{acknowledgements}


\begin{thebibliography}{99}

\bibitem{apr05} A. Aprahamian, K. Langanke, M. Wiescher, Prog. Part. Nucl. Phys. {\bf 54}, 535 (2005).

\bibitem{ang99} C. Angulo et al., Nucl. Phys. A {\bf 656}, 3 (1999).

\bibitem{nom85} K. Nomoto, F.-K. Thielemann, S. Miyaji, Astron. Astrophys. {\bf 149}, 239 (1985).

\bibitem{lan86}  K. Langanke, M. Wiescher, F.-K. Thielemann, Z. Phys. A  {\bf 324}, 147 (1986).

\bibitem{gar11} E. Garrido, R. de Diego, D.V. Fedorov, A.S. Jensen, Eur. Phys. J. A {\bf 47}, 102 (2011). 

\bibitem{tru69} J.W. Truran, B.Z. Kozlovsky, Astrophys. J. {\bf 158}, 10201 (1969).

\bibitem{mor70} J.F. Morgan, D.C. Weisser, Nucl. Phys. A {\bf 151}, 561 (1970).

\bibitem{dav71} C.N. Davids, T.I. Bonner, Astrophys. J. {\bf 166}, 405 (1971).

\bibitem{die10} R. de Diego, E. Garrido, D.V. Fedorov, A.S. Jensen, J. Phys. G: Nucl. Part. Phys. {\bf 37}, 115105 (2010). 

\bibitem{die14} R. de Diego, E. Garrido, D.V. Fedorov, A.S. Jensen, Eur. Phys. J. A {\bf 50}, 93 (2014). 

\bibitem{bea17} M. Beard, S.M. Austin, R. Cyburt, Phys. Rev. Lett. {\bf 119}, 112701 (2017).

\bibitem{fow67} W. A. Fowler, G. R. Caughlan, B. A. Zimmerman, Annu. Rev. Astron. Astrophys. {\bf 5}, 525 (1967).

\bibitem{gar14} E. Garrido, A. Kievsky, M. Viviani, Phys. Rev. C {\bf 90}, 014607 (2014).

\bibitem{gar18} E. Garrido, Few-Body Syst. {\bf 59}, 17 (2018).

\bibitem{gar05} E. Garrido, D.V. Fedorov, A.S. Jensen, and H.O.U. Fynbo, Nucl. Phys. A {\bf 748}, 27 (2005).

\bibitem{til04} D.R. Tilley, J.H. Kelley, J.L. Godwin, D.J. Millener, J. Purcell, C.G. Sheu, H.R. Weller, Nucl. Phys. A {\bf 745}, 155 (2004).

\bibitem{til02} D.R. Tilley, C.M. Cheves, J.L. Godwin, G.M. Hale, H.M. Hofmann, J.H. Kelley, C.G. Sheu, H.R. Weller, Nucl. Phys. A {\bf 708}, 3 (2002).

\bibitem{kel16} J.H. Kelley, J.E. Purcell, C.G. Sheu,  Nucl. Phys. A {\bf 968}, 71 (2016).

\bibitem{gib59} J.H. Gibbons, R.L. Macklin, Phys. Rev. {\bf 114}, 571 (1959).

\bibitem{gib65} J.H. Gibbons, R.L. Macklin, Phys. Rev. {\bf 137}, B1508 (1965).

\bibitem{ols89} N. Olsson, B. Trostell, E. Ramstr\"{o}m, Phys. Med. Biol. {\bf 34}, 909 (1989).


\end{thebibliography}
\end{document}